\documentclass[prd,aps,preprint,amsmath,nofootinbib,amssymb,eqsecnum,showkeys,tightenlines]{revtex4}
\pdfoutput=1
\usepackage{verbatim,graphics,graphicx,color,slashed}
\usepackage{ulem} 
\usepackage{hyperref}
\graphicspath{{figures/}}

\newcommand{\MeV}{\mathrm{MeV}}
\newcommand{\GeV}{\mathrm{GeV}}
\newcommand{\TeV}{\mathrm{TeV}}

\begin{document}

\title{Self-interacting scalar dark matter with local $Z_{3}$ symmetry}
\author{P. Ko and Yong Tang}
\affiliation{School of Physics, Korea Institute for Advanced Study,\\
 Seoul 130-722, Korea }
\date{\today}

\begin{abstract}
We construct a self-interacting scalar dark matter (DM) model with local discrete 
$Z_{3}$ symmetry that stabilizes  a weak scale scalar dark matter $X$.  
The model assumes a hidden sector with a local $U(1)_X$ dark gauge symmetry, 
which is broken spontaneously into $Z_3$ subgroup by nonzero VEV of dark Higgs 
field $\phi_X$ ($ \langle \phi_X \rangle \neq 0$).  Compared with global $Z_3$ DM models,  
the local $Z_3$ model has two new extra fields: a dark gauge field $Z^{'}$ and a dark Higgs 
field $\phi$ (a remnant of the $U(1)_X$ breaking).  After imposing various 
constraints including  the upper bounds on the spin-independent direct detection cross 
section and thermal relic density,   we find that the scalar DM  with mass less than 
$125$ GeV is allowed in the local $Z_3$  model, in contrary to the global $Z_3$ model.  
This is due to new channels in the DM pair annihilations open into $Z^{'}$ and $\phi$ 
in the local $Z_3$ model.  
Most parts of the newly open DM mass region can be probed by XENON1T and 
other similar future experiments.  
Also if $\phi$ is light enough  (a few MeV $\lesssim m_\phi \lesssim$ O(100) MeV), 
it can generate a right size of DM self-interaction and explain the astrophysical 
small scale structure anomalies. This would lead to exotic decays of 
Higgs boson into a pair of dark Higgs bosons, which could be tested at LHC and ILC.
\end{abstract}
\maketitle

\section{Introduction}
Although Planck \cite{Ade:2013zuv} has already given the dark matter(DM) relic density 
$\Omega h^2=0.1199\pm 0.0027$ with a high precision, we still do not know particle physics 
nature of DM at all.   So far all the compelling evidences for the existence of DM come from 
astrophysics and cosmology, due to its gravitational interaction. Still, many particle physics models for DMs have been proposed, and most of them have a stable collisionless cold 
DM(CCDM) candidate whose self-interaction can be ignored.

The collisionless cold DM has been very successful when explaining the large scale structure of our Universe. However, anomalies from the small scale astrophysical 
observations~\cite{Oh:2010ea,BoylanKolchin:2011de,BoylanKolchin:2011dk} indicate that 
DM may have strong interactions between themselves. Such self-interaction~\cite{Spergel:1999mh} would make DM have a flat core density profile rather than a cusp one predicted by CCDM. Recent simulations show that in order to flatten the cores of galaxies the cross section for DM scattering should be around $\sigma\sim  M_X \times \textrm{ barn }\GeV^{-1}$~\cite{Vogelsberger:2012ku, Rocha:2012jg, Zavala:2012us},  which is in fact a huge cross 
section compared with typical weak-scale cross sections $\sigma\sim 10^{-12}$ barn or 
$1$ pb.  Some light particle mediator in the dark sector could be an origin of such strong 
self-interaction between DMs.

In this paper, we propose a scalar DM model with a local $Z_3$ symmetry.  
Unlike models based on global symmetries, local discrete symmetries can protect 
symmetry-breaking from quantum gravity effects and guarantee the longevity or 
absolute stability of DM particles. 
Also a light mediator can exist in the models with local symmetry, and generate the correct 
self-interaction for DM in explaining the anomalies mentioned in  the previous paragraph. 

The outline of this paper is as follows.  In Sec.~II, we introduce the model with a local 
$Z_3$ symmetry, establish the convention for parameters and give the physical mass 
spectra.  Then we discuss both theoretical and experimental constraints on the parameters
in Sec.~III.  Then in Sec.~IV, we discuss the relic density and DM direct searches, 
paying attentions to the semi-annihilation feature, and compare with the global $Z_3$ 
mode.  In Sec.~V, we show that a light scalar mediator in our model can induce strong 
interaction for DM.  Finally we summarize the results in Sec.~VI. 

\section{Local $Z_3$ Model}

Let us assume the dark sector has a local $U(1)_{X}$ gauge which
is spontaneously broken into local $Z_{3}$ symmetry  a la Krauss and Wilczek~\cite{Krauss:1988zc} 
(see ref.~\cite{Batell:2010bp} for local $Z_N$ case). 
This can be achieved with two complex scalar fields 
\[
\phi_{X}\equiv\left(\phi_{R}+i\phi_{I}\right)/\sqrt{2}, 
 ~~~X\equiv\left(X_{R}+iX_{I}\right)/\sqrt{2}
 \] 
in the dark sector with the $U(1)_{X}$ charges equal to $1$ and $1/3$, respectively. 
Then one can write down renormalizable Lagrangian for the SM fields and the dark 
sector fields, $\tilde{X}_\mu, \phi_X$ and $X$: 
\begin{eqnarray}
{\cal L} & = & {\cal L}_{{\rm SM}}-\frac{1}{4}\tilde{X}_{\mu\nu}\tilde{X}^{\mu\nu}-\frac{1}{2}\sin\epsilon\tilde{X}_{\mu\nu}\tilde{B}^{\mu\nu}+D_{\mu}\phi_{X}^{\dagger}D^{\mu}\phi_{X}+D_{\mu}X^{\dagger}D^{\mu}X-V\nonumber \\
V & = & -\mu_{H}^{2}H^{\dagger}H+\lambda_{H}\left(H^{\dagger}H\right)^{2}-\mu_{\phi}^{2}\phi_{X}^{\dagger}\phi_{X}+\lambda_{\phi}\left(\phi_{X}^{\dagger}\phi_{X}\right)^{2}+\mu_{X}^{2}X^{\dagger}X+\lambda_{X}\left(X^{\dagger}X\right)^{2}\nonumber \\
 &  & {}+\lambda_{\phi H}\phi_{X}^{\dagger}\phi_{X}H^{\dagger}H+\lambda_{\phi X}X^{\dagger}X\phi_{X}^{\dagger}\phi_{X}+\lambda_{HX}X^{\dagger}XH^{\dagger}H+
 \left( \lambda_{3}X^{3}\phi_{X}^{\dagger}+H.c. \right)   \label{eq:potential}
\end{eqnarray}
where the covariant derivative associated with the gauge field $X^{\mu}$
is defined as $D_{\mu}\equiv\partial_{\mu}-i\tilde{g}_{X}Q_{X}\tilde{X}_{\mu}$. 
The coupling $\lambda_{3}$ can be chosen as real and positive since it is always possible 
to  redefine $X$ to absorb the phase.  

We are interested in the phase with the  following vacuum expectation values for the 
scalar fields in the model:
\begin{eqnarray}
\langle H\rangle=\frac{1}{\sqrt{2}}\left(\begin{array}{c}
0\\
v_{h}
\end{array}\right),\;\langle\phi_{X}\rangle=\frac{v_{\phi}}{\sqrt{2}},\;\langle X\rangle=0,\label{eq:vacuumstate}
\end{eqnarray}
where only $H$ and $\phi_{X}$ have non-zero vacuum expectation values(vev). This
vacuum will break electroweak symmetry into $U(1)_{\rm em}$, and $U(1)_{X}$ 
symmetry into local $Z_3$, which  stabilizes the scalar field $X$ and make it DM.
The discrete gauge $Z_3$ symmetry stabilizes the scalar DM even if we consider 
higher dimensional nonrenormalizable operators which are invariant under $U(1)_X$.
This is in sharp constrast with the global $Z_3$ model considered in Ref.~\cite{globalz3}.
Also the particle contents in local and global $Z_3$ models are different so that the 
resulting DM phenomenology are distinctly different from each other.
 
Other vacuum configurations could  exist, such as $\langle\phi_{X}\rangle\neq0$ and 
$\langle X\rangle\neq0$ which give rise to both broken $U(1)_{X}$ and $Z_{3}$ 
but also no dark matter candidate. The complete analysis of vacuum structure is
beyond the scope of this work and we shall focus on the vacuum 
Eq.~(\ref{eq:vacuumstate}) in this paper.

Expanding the scalar fields around Eq.~(\ref{eq:vacuumstate}), 
\begin{equation}
H\rightarrow\frac{v_{h}+h}{\sqrt{2}},\;\phi_{X}\rightarrow\frac{v_{\phi}+\phi}{\sqrt{2}},\; X\rightarrow\frac{x}{\sqrt{2}}e^{\mathrm{i}\theta}\textrm{ or }\frac{1}{\sqrt{2}}\left(X_{R}+iX_{I}\right),
\end{equation}
the minimum conditions for the potential would give %
\begin{eqnarray}
\left.\frac{\partial V}{\partial\phi}\right|_{x=0} & = & \phi\left(-\mu_{\phi}^{2}+\lambda_{\phi}\phi^{2}+\frac{1}{2}\lambda_{\phi H}h^{2}\right)=0 ,\\
\left.\frac{\partial V}{\partial h}\right|_{x=0} & = & h\left(-\mu_{H}^{2}+\lambda_{H}h^{2}+\frac{1}{2}\lambda_{\phi H}\phi^{2}\right)=0 .
\end{eqnarray}
Then one can solve them for the VEVs as follows: 
\begin{eqnarray}
\langle H^{2}\rangle & = & 
\frac{v_{h}^{2}}{2}=\frac{2\lambda_{\phi}\mu_{H}^{2}-\lambda_{\phi H}\mu_{\phi}^{2}}{4\lambda_{H}\lambda_{\phi}-\lambda_{\phi H}^{2}},
\\
\langle\phi_{X}^{2}\rangle & = & 
\frac{v_{\phi}^{2}}{2}=\frac{2\lambda_{H}\mu_{\phi}^{2}-\lambda_{\phi H}\mu_{H}^{2}}{4\lambda_{H}\lambda_{\phi}-\lambda_{\phi H}^{2}},
\end{eqnarray}
The mass matrix for the two mixed scalars is 
\begin{equation}
\mathcal{M}^{2}=\left(\begin{array}{cc}
2\lambda_{H}v_{h}^{2} & \lambda_{\phi H}v_{h}v_{\phi}\\
\lambda_{\phi H}v_{h}v_{\phi} & 2\lambda_{\phi}v_{\phi}^{2}
\end{array}\right)
\end{equation}
in the $(h, \phi)$ basis. 
Diagonalizing the mass matrix gives the mass eigenstates $H_{1}$
and $H_{2}$ 
\begin{equation}
\left(\begin{array}{c}
H_{1}\\
H_{2}
\end{array}\right)=\left(\begin{array}{cc}
\cos{\alpha} & {}-\sin{\alpha}\\
\sin{\alpha} & \cos{\alpha}
\end{array}\right)\left(\begin{array}{c}
h\\
\phi
\end{array}\right)
\end{equation}
and the mixing angle 
\[
\tan{2\alpha}=\frac{2\mathcal{M}_{12}^{2}}{\mathcal{M}_{22}^{2}-\mathcal{M}_{11}^{2}}=\frac{\lambda_{\phi H}v_{h}v_{\phi}}{\lambda_{\phi}v_{\phi}^{2}-\lambda_{H}v_{h}^{2}},\;\mathrm{or}\;\sin{2\alpha}=\frac{2\lambda_{\phi H}v_{h}v_{\phi}}{M_{H_{2}}^{2}-M_{H_{1}}^{2}}.
\]
Physical masses for $H_{1}$ and $H_{2}$ are 
\begin{equation}
M_{H_{1},H_{2}}^{2}=\lambda_{H}v_{h}^{2}+\lambda_{\phi}v_{\phi}^{2}\pm\sqrt{\left(\lambda_{H}v_{h}^{2}-\lambda_{\phi}v_{\phi}^{2}\right)^{2}+\left(\lambda_{\phi H}v_{h}v_{\phi}\right)^{2}}.
\end{equation}\label{eq:H1H2mass}
We shall identify $H_1$ as the recent discovered Higgs boson with $M_{H_1}\simeq 125\GeV$ and treat $M_{H_2}$ as a free parameter. $H_2$ could be either heavier or lighter than $H_1$.
The mass for the scalar DM $X$ is 
\[
M_{X}^{2}=\mu_{X}^{2}+\lambda_{\phi X}\frac{v_{\phi}^{2}}{2}
+\lambda_{HX}\frac{v_{h}^{2}}{2}.
\]

After the EW and dark gauge symmetry breaking, the mass terms for gauge fields are 
derived from 
\begin{equation}
\frac{v_{\phi}^{2}}{2}\tilde{g}_{X}^{2}\tilde{X}^{\mu}\tilde{X}_{\mu}+\frac{v_{h}^{2}}{8}\left(g_{1}\tilde{B}_{\mu}-g_{2}\tilde{W}_{3\mu}\right)^{2}.
\end{equation}
We can redefine the abelian gauge fields 
\begin{equation}
\left(\begin{array}{c}
\tilde{B}_{\mu}\\
\tilde{X}_{\mu}
\end{array}\right)=\left(\begin{array}{cc}
1 & -\tan\epsilon\\
0 & 1/\cos\epsilon
\end{array}\right)\left(\begin{array}{c}
\hat{B}_{\mu}\\
\hat{X}_{\mu}
\end{array}\right),\;\tilde{W}_{\mu}=\hat{W}_{\mu},
\end{equation}
in order to remove the kinetic mixing term between $\hat{B}_{\mu}$
and $\hat{X}_{\mu}$. We may also rescale the gauge coupling 
$\hat{g}_{X}=\tilde{g}_{X}/\cos\epsilon$. %
Substituting with the hatted field gives the mass matrix for $\hat{B}$,
$\hat{W}_{3}$ and $\hat{X}$, 
which we can diagonalize by rotating 
\begin{equation}
\left(\begin{array}{c}
\hat{B}_{\mu}\\
\hat{W}_{3\mu}\\
\hat{X}_{\mu}
\end{array}\right)=\left(\begin{array}{ccc}
c_{\tilde{W}} & -s_{\tilde{W}}c_{\xi} & s_{\tilde{W}}s_{\xi}\\
s_{\tilde{W}} & c_{\tilde{W}}c_{\xi} & -c_{\tilde{W}}s_{\xi}\\
0 & s_{\xi} & c_{\xi}
\end{array}\right)\left(\begin{array}{c}
A_{\mu}\\
Z_{\mu}\\
Z'_{\mu}
\end{array}\right).\label{eq:mixing1}
\end{equation}
Then the final mixing matrix for the starting fields in the lagrangian
is 
\begin{equation}
\left(\begin{array}{c}
\tilde{B}_{\mu}\\
\tilde{W}_{3\mu}\\
\tilde{X}_{\mu}
\end{array}\right)=\left(\begin{array}{ccc}
c_{\tilde{W}} & -\left(t_{\epsilon}s_{\xi}+s_{\tilde{W}}c_{\xi}\right) & s_{\tilde{W}}s_{\xi}-t_{\epsilon}c_{\xi}\\
s_{\tilde{W}} & c_{\tilde{W}}c_{\xi} & -c_{\tilde{W}}s_{\xi}\\
0 & s_{\xi}/c_{\epsilon} & c_{\xi}/c_{\epsilon}
\end{array}\right)\left(\begin{array}{c}
A_{\mu}\\
Z_{\mu}\\
Z'_{\mu}
\end{array}\right).\label{eq:mixing2}
\end{equation}
In Eq.~(\ref{eq:mixing1}) and (\ref{eq:mixing2}), we have defined the new parameters:  
\begin{eqnarray}
 &  & c_{\tilde{W}}\equiv\cos\theta_{\tilde{W}}=\frac{g_{2}}{\sqrt{g_{1}^{2}+g_{2}^{2}}},\;\tan2\xi=-\frac{m_{\tilde{Z}}^{2}s_{\tilde{W}}\sin2\epsilon}{m_{\tilde{X}}^{2}-m_{\tilde{Z}}^{2}\left(c_{\epsilon}^{2}-s_{\epsilon}^{2}s_{\tilde{W}}^{2}\right)},\nonumber \\
 &  & t_{x}\equiv\tan{x},\; c_{x}\equiv\cos{x}\;\mathrm{and}\; s_{x}\equiv\sin{x}\;\mathrm{for}\; x=\epsilon,\xi,\nonumber \\
 &  &m_{\tilde{X}}^{2}=\hat{g}_{X}^{2}v_{\phi}^{2},\; m_{\tilde{Z}}^{2}=\frac{1}{4}\left(g_{1}^{2}+g_{2}^{2}\right)v_{h}^{2}.
\end{eqnarray}
From Eq.~(\ref{eq:mixing2}) we can observe that the SM particles  charged under 
$SU(2)_{L}$ and/or $U(1)_{Y}$ now also have interaction with $Z'_{\mu}$.
And particles in the dark sector also have  interaction with $Z_{\mu}$ due to the kinetic 
mixing beween $\hat{B}_\mu$ and $\hat{X}_\mu$. 

The physical masses for four vector bosons in our model are given by 
\begin{eqnarray}
m_{A}^{2}& = & 0,   \\
m_{W}^{2} & = & m_{\tilde{W}}^{2}=\frac{1}{4}g_{2}^{2}v_{h}^{2},  
\\ m_{Z}^{2} & = & m_{\tilde{Z}}^{2}\left(1+s_{\tilde{W}}t_{\xi}t_{\epsilon}\right),
\\ 
m_{Z'}^{2} & = & \frac{m_{\tilde{X}}^{2}}{c_{\epsilon}^{2}\left(1+s_{\tilde{W}}t_{\xi}t_{\epsilon}\right)}.
\end{eqnarray}

\section{Constraints on $\lambda$'s and $\epsilon$}
The dimensionless parameters $\lambda_i$'s can not be arbitrarily large in  perturbative 
theory. Demanding $|\lambda_i|\lesssim 4\pi$ would be sufficient for our consideration. 
The scale where perturbativity breaks down can be determined by  the renormalization 
group (RG) analysis using the RG equations summarized in Appendix. 
Generally,  $|\lambda_i|\lesssim 1$ at the TeV scale would give perturbativity up to 
$10^{15}$ GeV.

Besides the perturbativity, the potential should be bounded from below, which means at large field
value the potential needs to be positive semidefinite, $V\geq0$. In
the limit of $\phi_{i}\rightarrow\infty$, we can neglect the quadratic
terms and consider only the quartic part in the potential. Then in
the case of $\lambda_{3}=0$ we have~\cite{copositivity1,copositivity2,Kannike:2012pe} 
\begin{eqnarray}
 &  & \lambda_{H}\geq0,\;\lambda_{\phi}\geq0,\;\lambda_{X}\geq0,\; A_{\phi H}\equiv\lambda_{\phi H}+2\sqrt{\lambda_{\phi}\lambda_{H}}\geq0,\;\label{eq:lambdaconstraint1}\\
 &  & A_{\phi X}\equiv\lambda_{\phi X}+2\sqrt{\lambda_{\phi}\lambda_{X}}\geq0,\; A_{HX}\equiv\lambda_{HX}+2\sqrt{\lambda_{H}\lambda_{X}}\geq0,\;\\
 &  & \sqrt{\lambda_{H}\lambda_{\phi}\lambda_{X}}+\lambda_{\phi H}\sqrt{\lambda_{X}}+\lambda_{\phi X}\sqrt{\lambda_{H}}+\lambda_{HX}\sqrt{\lambda_{\phi}}+\sqrt{A_{\phi H}A_{\phi X}A_{HX}}\geq0.
\end{eqnarray}
For general $\lambda_{3}$, there is no transparent criteria for the
positive semidefinite of $V_{4}$. However we could get useful necessary
conditions by using the general positive criteria for quartic polynomial. For
instance, consider the direction in the field space, 
\[
h=0,\;\phi_{X}=y\times x,\; X=\frac{x}{\sqrt{2}},
\]
and substitute in the quartic potential, we have 
\[
V_{4}=\frac{1}{4}\left(\lambda_{\phi}y^{4}+\lambda_{\phi X}y^{2}+2\lambda_{3}y+\lambda_{X}\right)x^{4}.
\]
Boundness from below gives the constraints on the coefficients for
any non-negative $y$ 
\[
\lambda_{\phi}y^{4}+\lambda_{\phi X}y^{2}+2\lambda_{3}y+\lambda_{X}\geq0,
\]
of which one sufficient condition is 
\begin{align*}
& 0 <\beta\equiv\frac{\lambda_{\phi X}}{\sqrt{\lambda_{\phi}\lambda_{X}}}\leq6\textrm{ \&\& }\gamma\equiv\frac{2\lambda_{3}}{\left(\lambda_{\phi}\lambda_{X}^{3}\right)^{\frac{1}{4}}}>-\frac{\beta+2}{2},\\
\textrm{or } & \beta>6\textrm{ \&\& }\gamma>-2\sqrt{\beta-2}.
\end{align*}
General conditions for positivity on a quartic polynomial of a single variable~\cite{polynomial} is summarized 
in the Appendix B and shall be imposed in all following investigations. 

Similarly we can do the analysis in another  directions. 
The direction $h=y\times x,\;\phi_{X}=0,\; X=\dfrac{x}{\sqrt{2}}$  gives 
\[
\lambda_{H}y^{4}+\lambda_{HX}y^{2}+\lambda_{X}\geq0,
\]
leading constraints on $\lambda_{H}$, $\lambda_{X}$ and $\lambda_{HX}$
which are just those in Eq. (\ref{eq:lambdaconstraint1}).  

Constraints on the kinetic mixing parameter $\epsilon$ come from the muon $(g-2)$, 
atomic parity  violation, the $\rho$ parameter and electroweak precision tests(EWPTs)~\cite{Fayet:2007ua, Chun:2010ve, Kumar:2006gm, Chang:2006fp}. 
These could put an upper limit on $\epsilon$ as a function of $M_{Z'}$. 
Among  these constraints, EWPTs provides the most stringent one:  
\begin{equation}
\left(\frac{\tan{\epsilon}}{0.1}\right)^2\left(\frac{250\GeV}{M_{Z'}}\right)^2\leq0.1.
\label{eq:EWPTs}
\end{equation} 
For $M_{Z'}\sim 250\GeV$ we have $\epsilon\lesssim 0.03$. In the case of $\epsilon=0$, there is no mixing
between $Z$ and $Z'$, the whole connnection between SM and dark
sector comes from the scalar sector.

In the following numerical investigation, we have imposed all the relevant constraints discussed in this section.

\section{Relic density and direct detection}

\begin{figure}
\includegraphics[width=1\textwidth]{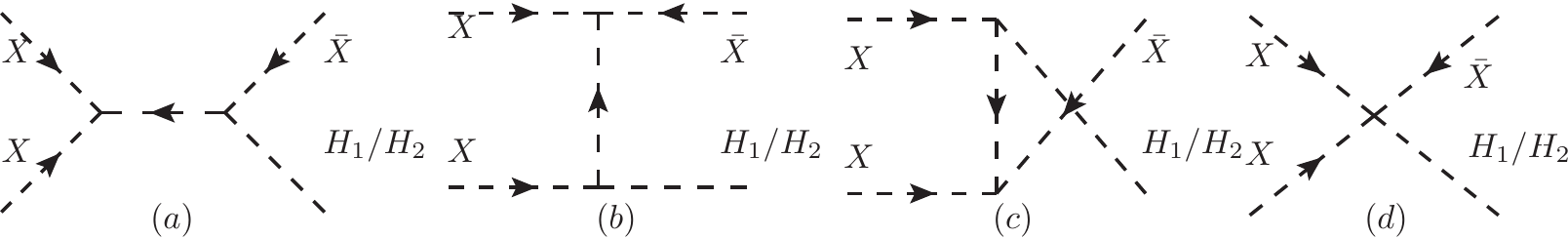}
\includegraphics[width=0.8\textwidth]{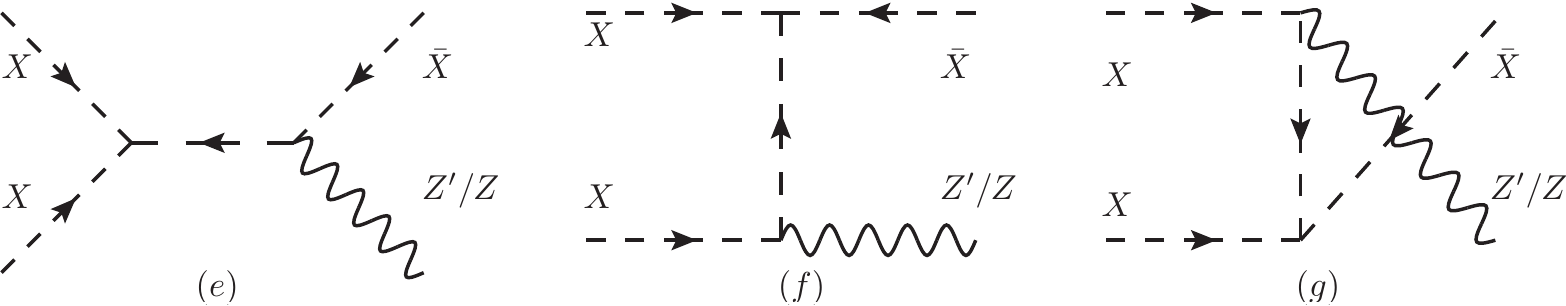} 
\caption{Feynman diagrams for dark matter semi-annihilation. Only (a), (b), and (c) with $H_1$ as final state appear in the global $Z_3$ model, while all diagrams could contribute in local $Z_3$ model.
\label{fig:semi-annihilation}}
\end{figure}

\begin{figure}
\includegraphics[width=1\textwidth]{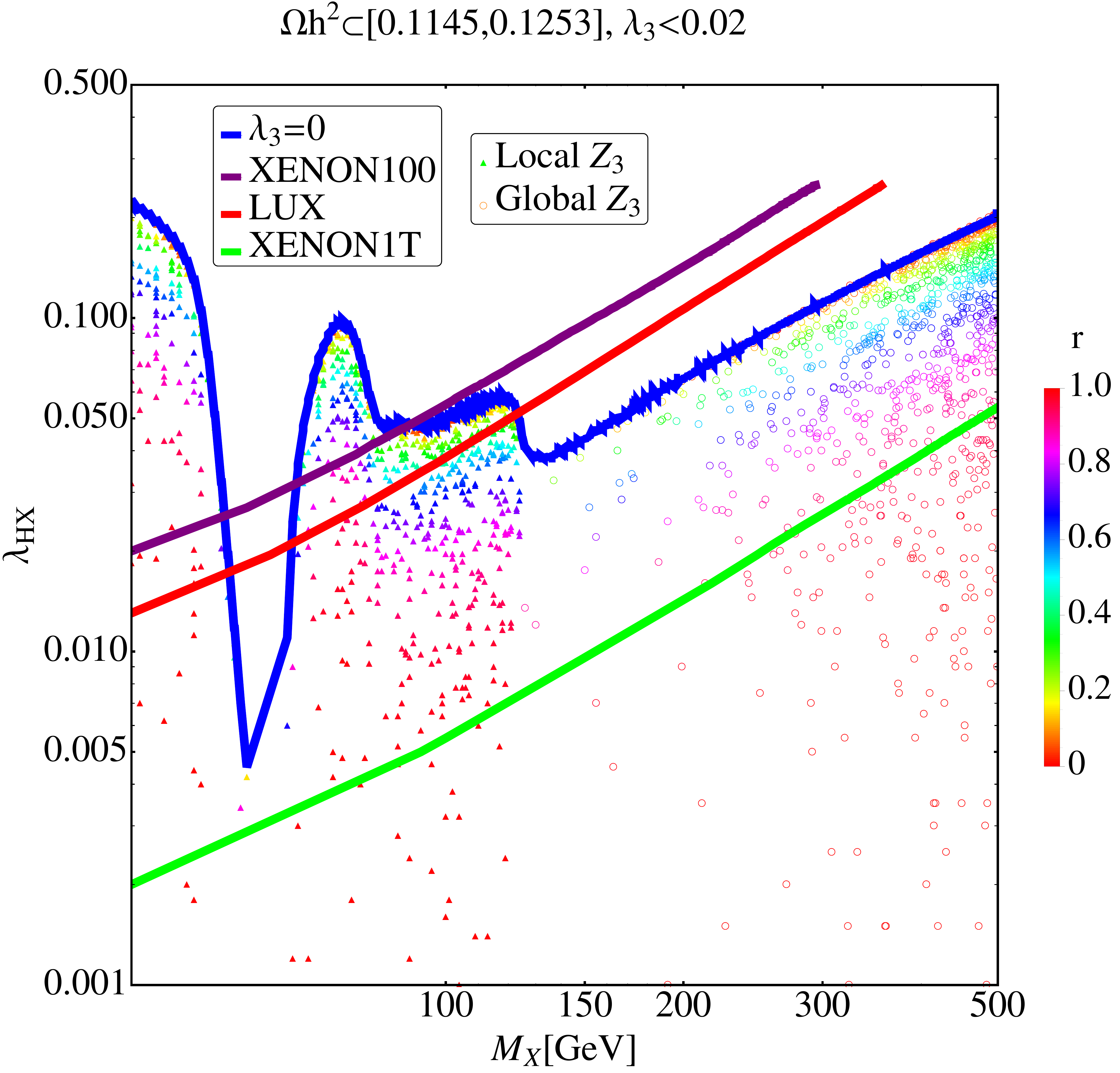}
\caption{Illustration of discrimination between global and local $Z_3$ symmetry. We have chosen $M_{H_2}=20\GeV,\;M_{Z'}=1\TeV,\;\lambda_3<0.02,\; \epsilon\simeq 0$ and $\lambda_{\phi H}\simeq 0$ as an example. From up to down, three nearly straight lines 
mark the XENON100~\cite{xenon100}, LUX~\cite{lux} and expected XENON1T limits~\cite{xenon1t}, respectively. Colors in the scatterred triangles and circles indicate the relative contribution of semi-annihilation, $r$. The curved blue band, together with the cirles, gives correct relic density of $X$ in the global $Z_3$ model. And the colored triangles appears 
only in the local $Z_3$ model. See text for detail.
\label{fig:global_gauge}}
\end{figure}

\subsection{semi-annihilation}

The $X^3\phi_X$ term and the cubic term $X^{3}$ after $U(1)_X$ symmetry breaking 
lend the semi-annihilation channel possible and could have a significant  effect 
in the freeze out of the DM~\cite{Hambye:2008bq,D'Eramo:2010ep,Belanger:2012vp}. 
We show the relevant Feynman diagrams in Fig.~\ref{fig:semi-annihilation}. 
In the presence of semi-annihilation the Boltzman equation that determines the number density $n_X$ is modified into~\cite{micromegas}
\begin{equation}
\frac{dn_X}{dt}=-v\sigma^{XX^{\ast}\rightarrow YY}\left(n_X^{2}-n_{X\textrm{ eq}}^{2}\right)-\frac{1}{2}v\sigma^{XX\rightarrow X^{\ast}Y}\left(n_X^{2}-n_Xn_{X\textrm{ eq}}\right)-3Hn_X,
\end{equation}
where $Y$ stands for any other particles and $v$ for the relative
velocity. Due to the semi-annihilation, new contribution appears
as the second term in the above equation. The numerical investigation is done 
with \texttt{micrOMEGAs}~\cite{micromegas}. 
We may define the fraction of the contribution from the semi-annihilation in terms of  
\[
r\equiv\frac{1}{2}\frac{v\sigma^{XX\rightarrow X^{\ast}Y}}{v\sigma^{XX^{\ast}\rightarrow YY}+\frac{1}{2}v\sigma^{XX\rightarrow X^{\ast}Y}}.
\]

The full Feynman diagrams for semi-annihilation are presented in 
Fig.~\ref{fig:semi-annihilation}. 
Depending  on the particles' masses or couplings, only a fraction of these diagrams 
might be kinematically allowed or relevant. For example, only first four diagram are 
relevant for $\epsilon\simeq0,\;\lambda_{\phi X}\simeq 0$ and very heavy $Z'$. 
Then the cross section for  $XX\rightarrow X^\ast H_i$ 
semi-annihilation process is
\[
\frac{d\sigma}{d\Omega}=\frac{1}{64\pi^{2}s}\frac{\left|p_{f}\right|}{\left|p_{i}\right|}\left|\mathcal{M}\right|^{2},
\]
with $\left|p_{f}\right|=\dfrac{1}{2\sqrt{s}}\sqrt{\left[s-\left(M_{X}+M_{H_i}\right)^{2}\right]\left[s-\left(M_{X}-M_{H_i}\right)^{2}\right]}$. For
dark matter $p_{i}=M_{X}v_{\textrm{vel}}/2$ and $v_{\textrm{vel}}$ is the relative velocity between
two annihilating particles. Matrix elements are given by 
\begin{align*}
i \mathcal{M}_{d} & \propto-i3\sqrt{2}\lambda_{3},\\
i \mathcal{M}_{a+b+c} & \propto-i3\sqrt{2}\lambda_{3}v_{\phi}\left[\frac{i}{s-M_{X}^{2}}
+\frac{i}{t-M_{X}^{2}}+\frac{i}{u-M_{X}^{2}}\right]\left(-i\lambda_{HX}v_{h}\right) ,
\end{align*}
respectively. 
If $\lambda_{HX}v_{h}v_{\phi}/M_{X}^{2}\ll1$ and $M_{H_{i}}<M_{X}$, then 
$\mathcal{M}_{d}$ dominates and we have 
\[
\left\langle \sigma v\right\rangle _{d}=\frac{9\lambda_{3}^{2}}{16\pi}\frac{\left|p_{f}\right|}{M_{X}^{3}},\; \textrm { and }\left|p_{f}\right|\simeq\dfrac{3}{4}M_{X} \textrm{ for }M_{X}\gg M_{H_i}.
\]
The relevant contribution  $r$  from semi-annihilation is shown with different color in Fig.~\ref{fig:global_gauge}. 
It is evident that as $\lambda_{HX}$ gets smaller, $r$ becomes larger and 
the semi-annihilation becomes dominant.  Meanwhile the cross section for $X$'s scattering 
off a nucleon gets smaller for direct searches. Some of these points may even not be 
probed by XENON1T~\cite{xenon1t}.

\subsection{Global $Z_3$ vs Local $Z_3$}
When the $U(1)_{X}$ breaking scale $v_{\phi}$ is much larger than
the EW scale $v_{h}$ and the masses, $M_{Z'}$ and $M_{H_2}$, 
are much heavier than those of other particles, we can get the low energy effective
theory by integrating out the heavy degrees of freedom, $X^{\mu}$
and $\phi$. The effective theory then describes the SM$+X$ with
the residual global $Z_{3}$ symmetry. And in the effective potential the
terms involving $X$ always appears as $X^{\dagger}X$, $X^{3}$
and $X^{\dagger3}$, 
\begin{align}
V_{\textrm{eff}}  \simeq  &-\mu_{H}^{2}H^{\dagger}H+\lambda_{H}\left(H^{\dagger}H\right)^{2}+\mu_{X}^{2}X^{\dagger}X+\lambda_{X}\left(X^{\dagger}X\right)^{2} + \lambda_{HX}X^{\dagger}XH^{\dagger}H +\mu_{3}X^{3} \nonumber\\
&+\textrm{higher order terms}+H.c,\label{eq:globalpotential}
\end{align}
where $\mu_3\equiv\lambda_3\dfrac{v_\phi}{\sqrt{2}}$. In such a case, the effective theory 
can not tell whether the $Z_{3}$ symmetry is a global one or just residual of a gauge 
symmetry. In fact the renormalizable parts of $V_{\rm eft}$ in Eq. (4.2) is exactly the same
as the scalar potential in global $Z_3$ model~\cite{globalz3}.  Therefore we can consider the renormalizable
scalar DM model with global $Z_3$ symmetry as an effective theory of local $Z_3$ models 
in the limit $v_\phi >> v_h$.  

However there is an important difference in the higher dimensional operators 
even in this limit. Within the local $Z_3$ model, the discrete $Z_3$ 
gauge symmetry is respected by higher dimenionsional operators, and the scalar DM $X$ 
shall be absolutely stable.  This is not the case for global $Z_3$ model, since the higher 
dimensional operators due to quantum gravity could break global $Z_3$ symmetry, so that
the DM stability is no longer guaranteed.  For example one can consider 
\[
\frac{1}{\Lambda} X F_{\mu\nu} F^{\mu\nu} \ ,  
\]
which renders the scalar $X$ with EW scale mass decay immediately, and so the 
scalar $X$ cannot make a good DM candidate of the universe. 

The difference between local and global $Z_3$ models become even more apparent and 
significant  when $v_{\phi}\sim \TeV$ or smaller.
There is only one additional new particle $X$ in the global $Z_3$ model, while in the local $Z_3$ model there are two more particles,  $Z'$ and $H_2$, compared with the global 
$Z_3$ model.  The particle spectra are different,  and the local $Z_3$ model enjoys 
much richer phenomenology. In Fig.~\ref{fig:global_gauge} 
we show an example that could illustrate the differences between the global and local 
$Z_3$ models. For simplicity we use $M_{H_2}=20\GeV,\;M_{Z'}=1\TeV,\;\lambda_3<0.02,\; \epsilon\simeq 0$ and $\lambda_{\phi H}\simeq 0$. 
The curved blue band shows the parameter region in which only 
$X X^\ast \rightarrow \textrm{SM+SM}$ processes contribute to annihilation, namely, only 
$\lambda_{HX}X^{\dagger}XH^{\dagger}H$ in the potential is relevant and it also marks 
the upper bound for $\lambda_{HX}$ for giving the correct relic abundance of $X$ 
in both global and local $Z_3$ models.
We can see that the low mass range $M_X<M_{H_1}$ is excluded by latest dark matter 
direct search limit from LUX~\cite{lux}, except the resonance region $M_X\simeq M_{H_1}/2$
which will be probed by XENON1T~\cite{xenon1t}.  Colored circles, together with the very 
curved blue band, describe the parameter space for the global $Z_3$ model where 
$X^{3}$-term comes to play since semi-annhilation happens here only when 
$M_X>M_{H_1}$. However, unlike the global model, local $Z_3$ model allows ample 
parameter space in the low mass range, $M_X<M_{H_1}$, even if LUX limit is taken into 
account. This is shown as colored triangles in Fig.~\ref{fig:global_gauge}. 

There could exist other differences between local and global $Z_3$ models. 
Depending on the exact value of $M_{Z'}$, $M_{H_2}$ and other physical parameters, 
the phenomena could be quite different.  For instance, when $Z'$ or $H_2$ is light, 
$H_1$ can decay to them if $\epsilon\neq 0$ or $\lambda_{\phi H}\neq 0$ (see Ref.~
\cite{Curtin:2013fra} for extensive survey and Ref.~\cite{Chpoi:2013wga}  
for the comprehensive study of a singlet scalar ($\phi$) 
mixing  with the SM Higgs boson). Also, in local $Z_3$ model isospin-violating interaction between DM and nucleon can arise from $Z^{'}$ exchange.
On the other hand,  only isospin-conserving couplings between DM and nucleon exist 
in global $Z_3$ model through the Higgs mediation, if we neglect small isospin violation 
from $m_u \neq m_d$.   Therefore one can have two independent channels in the 
DM-nucleon  scattering amplitude, which might be helpful to understand the recent data on 
direct detection of DM in the light WIMP region~\cite{Belanger:2013tla}. 
This is generic in models with local dark gauge symmetry which is spontaneously broken 
by dark Higgs field~\cite{preparation}. 

Finally, when $M_{Z'}$ or/and $M_{H_2}$ is about $\mathcal{O}(\MeV)$, sizable DM 
self-interaction could be realized, which is motivated to solve the astrophysical small scale 
structure anomalies. We shall discuss this self-interacting DM scenario in 
Sec.~\ref{sec:self} in detail. 

\subsection{Comparison with the effective field theory (EFT) approach}

In this subsection, we make a brief comparison of the renormalizable local $Z_3$ scalar 
DM model with the effective field theory (EFT) approach.  Usual starting point for the EFT 
approach is to write down the operators for direct detections of DMs.  For a complex 
scalar DM $X$ we are considering in  this work, one can easily construct the following 
operators imposing $Z_3$ symmetry,  to list only a few:
\begin{eqnarray}
U(1)_X ~{\rm sym} : && X^\dagger X H^\dagger H ,  \  
\frac{1}{\Lambda^2} \left( X^\dagger D_\mu X \right) 
\left( H^\dagger D^\mu H \right) , \  
\frac{1}{\Lambda^2} \left( X^\dagger D_\mu X \right) 
\left( \overline{f} \gamma^\mu f \right) ,  \ etc. 
\\
Z_3~ {\rm sym} : && 
\frac{1}{\Lambda} X^3 H^\dagger H , \ \ \frac{1}{\Lambda^2} X^3  \overline{f} f , \ etc.
\\
&& ({\rm or}~ \frac{1}{\Lambda^3} X^3  \overline{f_L} H f_R  , 
{\rm if~ we ~imposed ~the ~full~ SM ~gauge~ symmetry)}
\end{eqnarray}
where $f$ is a SM fermion field and $\Lambda$ is a combination of new physics scale
and couplings of the DM particle to new physics particle, and can differ from one operator
to another. The usual story within the EFT is that the direct detection cross section due to 
the renormalizable operator $X^\dagger X H^\dagger H$ is strongly constrained so that  
the scalar DM can not be thermalized if it is light.

Note that within the EFT picture there is no room for $Z^{'}$ or $H_2 (\approx 
\phi)$ to enter and play important roles in direct and indirect detection or in the calculation 
of DM thermal relic density. This is because we do not know which fields are relevant 
(or dynamical) at the energy scale we are considering.  Without constructing a full 
theory which is mathematically consistent and physically sensible, it would be difficult to
guess which fields would be relevant beforehand within the EFT approach.

Also note that the usual complementarity does not work in this $Z_3$ models, since the 
EFT approach for direct detection based on Eq.~(4.3) does not capture 
the semi-annihilation channels for thermal relic density or indirect DM signatures described
by Eqs.~(4.4) and (4.5),  which is unique in the $Z_3$ models. 
This simple example shows that the DM EFT can be useful only if we know the detailed 
quantum numbers of DM particle, such as its spin and other (conserved) quantum 
numbers. Otherwise the complementarity does not work.  Since we do not know anything
about the DM quantum numbers as of now, the EFT approach and complementarity 
arguments should be taken with a great caution. Otherwise one would make erroneous 
conclusions. 

More detailed discussions on the subtleties and limitations of EFT approach for 
DM physics will be  discussed elsewhere~\cite{progress}.

\section{Self-Interacting Dark Matter $X$}\label{sec:self}

\begin{figure}[thb]
\includegraphics[width=0.48\textwidth]{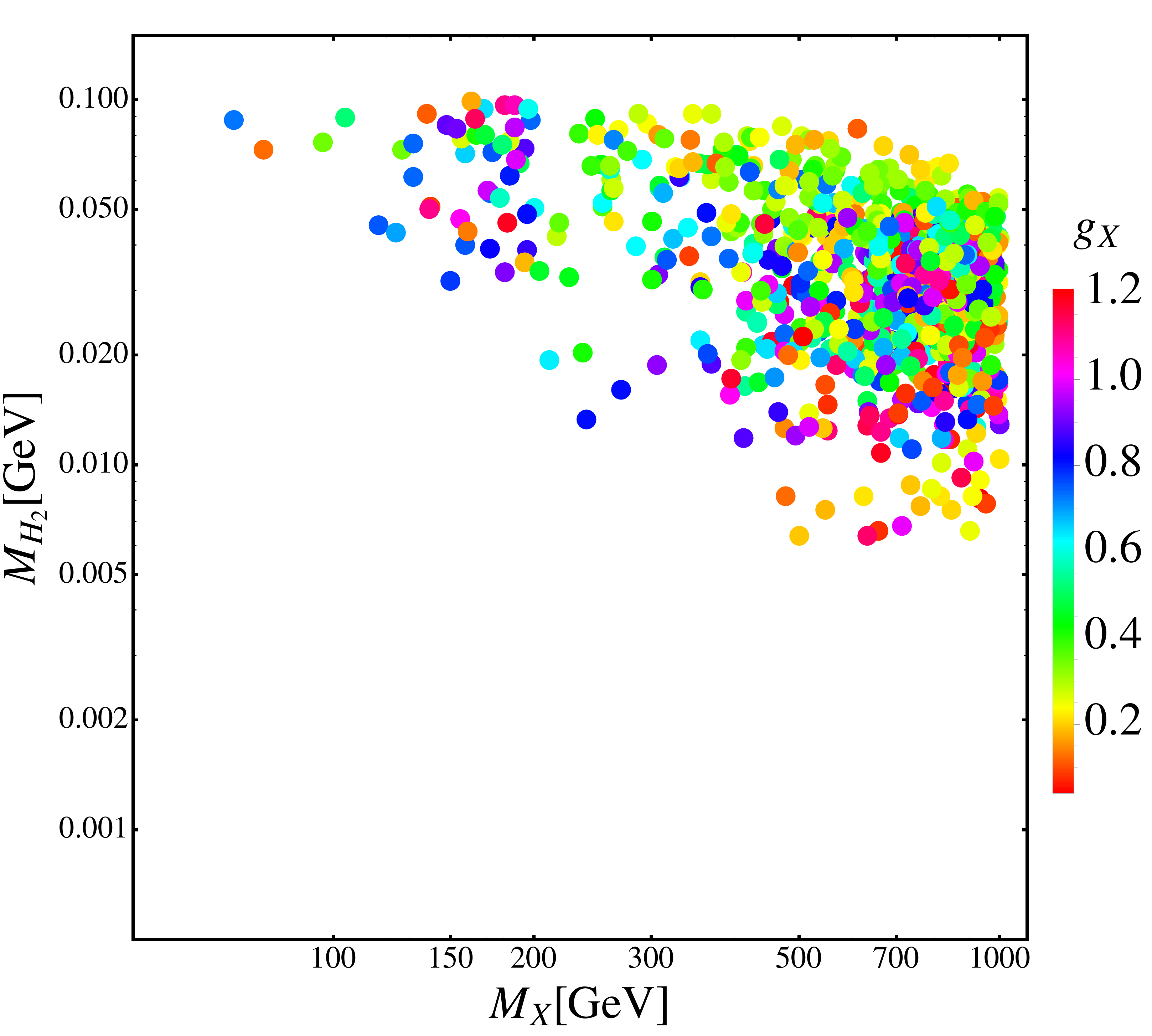}
\includegraphics[width=0.48\textwidth]{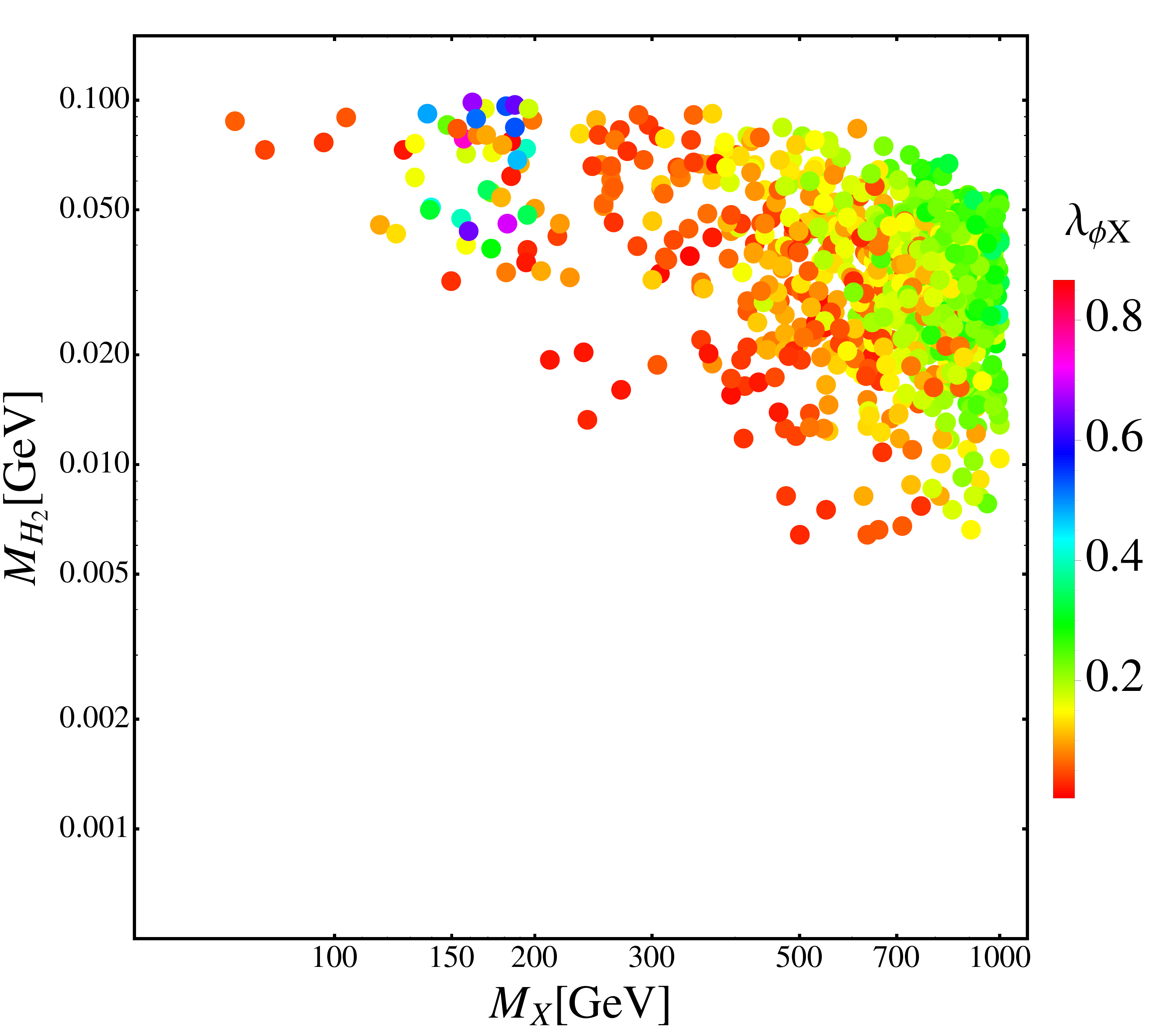}
\includegraphics[width=0.48\textwidth]{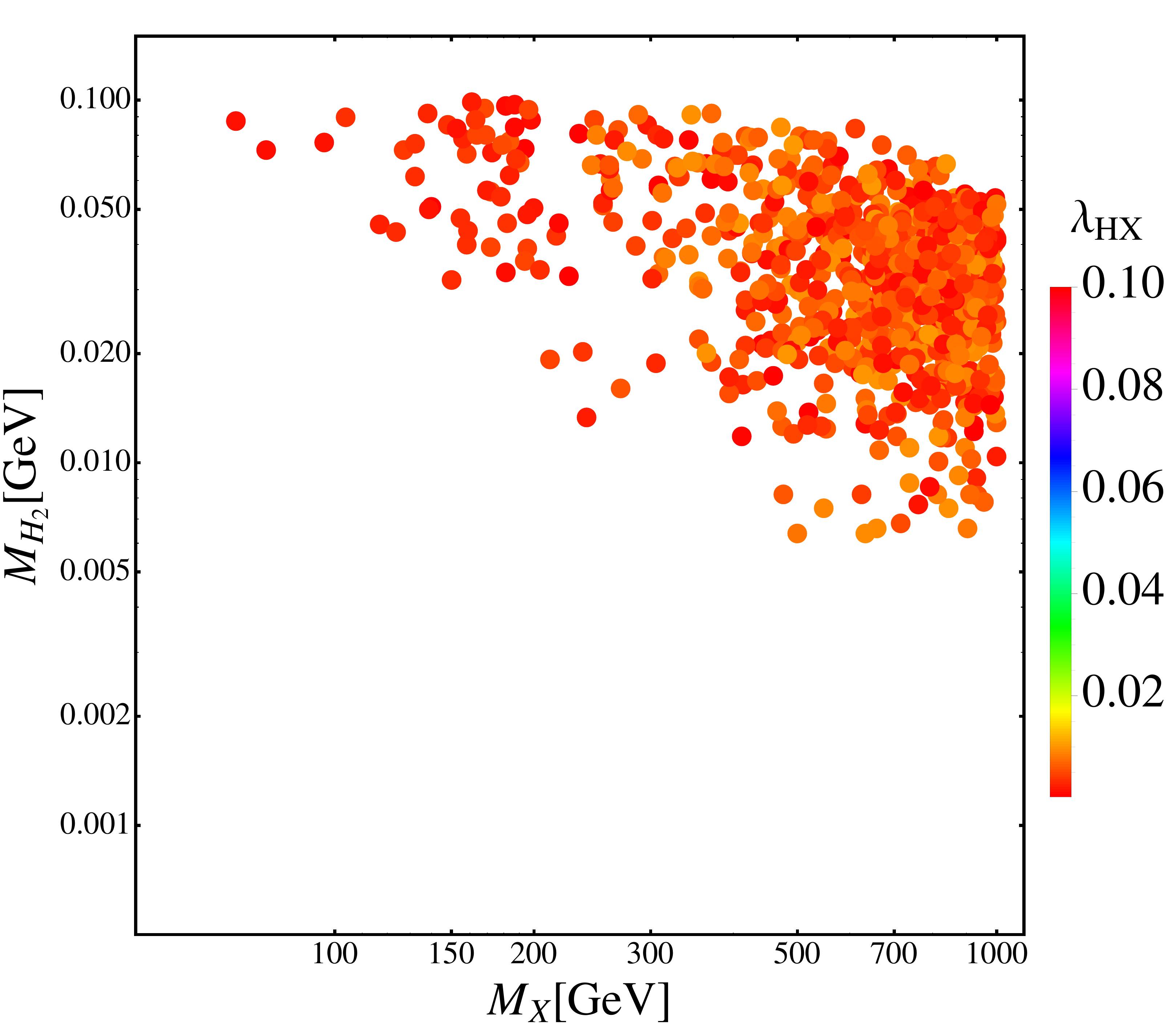}
\includegraphics[width=0.48\textwidth]{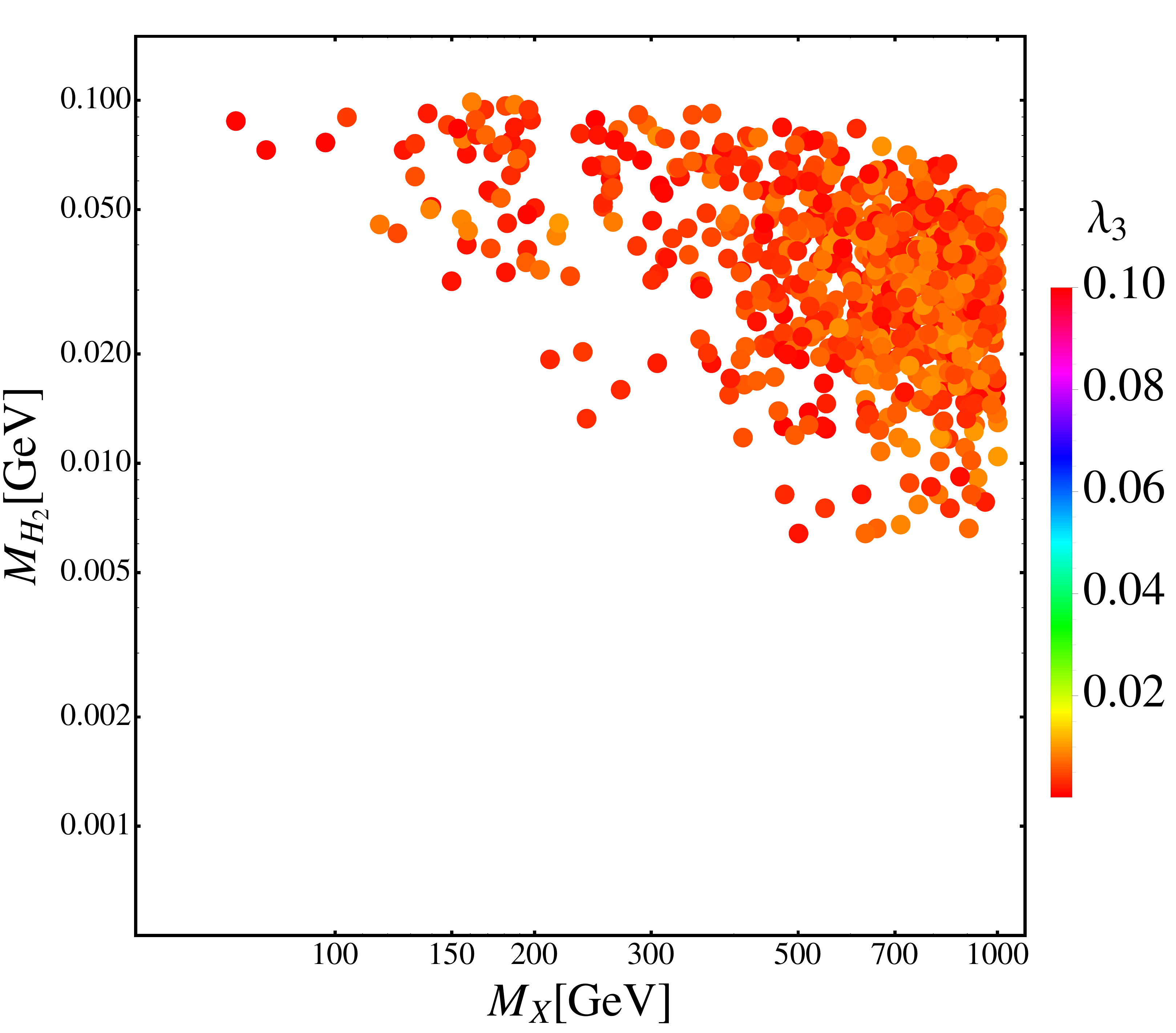}
\caption{Scatter plots of various parameters that are consistent with relic
density, LUX direct search bound and self-interaction $\sigma_{T}/M_{X}\in\left[0.1,10\right]\textrm{ cm}^{2}/\textrm{g}$ at Dwarf galaxies scale with $v_\textrm{rel}\simeq 10$ km/s, and $\sigma_{T}/M_{X}\lesssim 0.5\textrm{ cm}^{2}/\textrm{g}$ at Milky Way and cluster scales with $v_\textrm{rel}\simeq 220$ km/s and $v_\textrm{rel}\simeq 1000$ km/s, respectively. We have used $M_{Z'}\simeq 200\GeV$ and $\epsilon\ll0.03$ and scanned other parameters as illustration. \label{fig:SIDM} }
\end{figure}

One more difference between local and global $Z_3$ models is that there can exist strong 
self-interaction between scalar DM $X$ in the local $Z_3$ model~\footnote{This feature is 
not unique to local $Z_3$ model, but could appear in many other DM models with dark 
gauge symmetries. Another example with local $Z_2$ symmetry will be presented elsewhere
~\cite{preparation}.}. 
Traditional collisionless 
cold dark matter(CDM) can explain the large scale structure of the Universe. 
However, astrophysical anomalies in small scale structures motivate collisional CDM, 
which has self-interaction around $\sigma/M_{X}\sim0.1-10\textrm{ cm}^{2}\textrm{/g}$. 
This can be achieved in the local $Z_3$ model with $\mathcal{O}(\MeV)$ $H_2$ or $Z'$. 
A vector $Z'$ can mediate both attractive and replusive forces, and has been considered 
in~\cite{Feng:2009hw, Buckley:2009in, Loeb:2010gj, Aarssen:2012fx, Tulin:2012wi, Tulin:2013teo, Hannestad:2013ana, Dasgupta:2013zpn, Bringmann:2013vra}. 
So here we shall only concentrate on the $\mathcal{O}(\MeV)$ $H_2$ case in which only attractive 
force is mediated for explanation of small scale structures.  Other different phenomenologies 
of a light mediator can be found in~\cite{Pospelov:2007mp, Hooper:2008im, Feng:2008ya, Feng:2008mu, ArkaniHamed:2008qn, Pospelov:2008jd, Shepherd:2009sa, Kaplan:2009ag, 
An:2009vq}. 
 
Consider the $X X^\ast \rightarrow X X^\ast$ elastic scattering process
mediated by a t-channel scalar $H_{2}$, the differential
cross section is
\[
\frac{d\sigma}{d\Omega}=\frac{1}{64\pi^{2}s}\frac{\left|p_{f}\right|}{\left|p_{i}\right|}\left|\mathcal{M}\right|^{2},\;
\mathcal{M}\propto \frac{\lambda_{\phi X}^{2}v_{\phi}^{2}}{\left(p_{1}-p_{3}\right)^{2}-M_{H_2}^{2}},
\]
\[
\left(p_{1}-p_{3}\right)^{2}=2M_{X}^{2}-2\left(E_{1}E_{3}-\vec{p}_{1}\cdot\vec{p}_{3}\right)=-2\left|\vec{p}_{1}\right|^{2}\left(1-\cos\theta\right).
\]
Since $\left|p_{f}\right|=\left|p_{i}\right|,\; s\simeq4M_{X}^{2}$, $E_{1}=E_{3}$ and $\left|\vec{p}_{1}\right|=\left|\vec{p}_{3}\right|$
in the centre-of-mass system, then we have 
\begin{align}
\sigma_{\mathrm{SI}}=\int d\Omega\frac{d\sigma}{d\Omega} =\frac{\lambda_{\phi X}^{4}v_{\phi}^{4}}{64\pi M_{X}^{2}}\frac{1}{M_{H_2}^{2}\left(4\left|\vec{p}_{1}\right|^{2}+M_{H_2}^{2}\right)}\simeq\frac{\lambda_{\phi X}^{4}v_{\phi}^{4}}{64\pi M_{X}^{2}}\frac{1}{M_{H_2}^{4}},\textrm{ for }\left|\vec{p}_{1}\right|\ll M_{H_2}.\label{eq:SIcs}
\end{align}
The more relevant quantity for quantifying the self-interaction of DMs is the
momentum-transfer or transport cross section\footnote{
If the scattering particles are identical, $XX\rightarrow XX$ for instance, it may be more appropriate
to use the $\sigma_{T}\equiv\int d\Omega\left(1-\cos^{2}\theta\right)\dfrac{d\sigma}{d\Omega}$ which regularizes both forward and backward scattering~\cite{Cline:2013pca}(see similar discussion in \cite{Tulin:2013teo}).}
\[
\sigma_{T}\equiv\int d\Omega\left(1-\cos\theta\right)\frac{d\sigma}{d\Omega},
\]
which regularizes the forward scattering($\theta=0$) at which no momentum is transfered. 
In our case, we have for $XX^*\rightarrow XX^*$ scattering
\begin{equation}
\sigma_{T}=\frac{\lambda_{\phi X}^{4}v_{\phi}^{4}}{32\pi M_{X}^{2}}\left(\frac{1}{4\left|\vec{p}_{1}\right|^{2}}\right)^{2}\left[\ln\left(1+R^{2}\right)-\frac{R^{2}}{1+R^{2}}\right],\;\textrm{where }R^{2}=\frac{4\left|\vec{p}_{1}\right|^{2}}{M_{H_2}^{2}}.\label{eq:transfercs}
\end{equation}
This formula is consistent with \cite{Feng:2009hw} where a vector mediator is considered. We may rewrite the above equation as
\[ 
\sigma_{T}=\frac{2\pi}{M_{H_2}^{2}}\beta^{2}\left[\ln\left(1+R^{2}\right)-\frac{R^{2}}{1+R^{2}}\right],
\textrm{ where } \alpha_{\phi}\equiv\frac{\lambda_{\phi X}^{2}}{4\pi}\left(\frac{v_{\phi}}{2M_{X}}\right)^{2}
\textrm{ and } \beta\equiv\frac{2\alpha_{\phi}M_{H_2}}{M_{X}v_{\textrm{rel}}^{2}}.
\]
On the other hand, annihilation cross section for $X X^\ast \rightarrow\phi\phi$
at the freezing out time is approximately
\[
\sigma_{\mathrm{ann}}\simeq\frac{\lambda_{\phi X}^{4}v_{\phi}^{4}}{64\pi M_{X}^{2}}\frac{3}{M_{X}^{4}},
\]
which is much suppressed by $M^4_{H_2}/M^4_X$, compared with Eq.s~(\ref{eq:SIcs}) 
and (\ref{eq:transfercs}). Naive estimates suffice to show that if we have 
$\sigma_{\textrm{ann}}\sim O(1)$ pb for $M_X\sim O(1) \GeV$, then 
$M_{H_2}\sim O(1)-O(100) \MeV$ would give $\sigma_{\textrm{SI}}\sim O(1)$ barn and 
$\sigma_{T}/M_{X}\sim 1\textrm{ cm}^{2}/\textrm{g}$, 
although more delicated analysis would involve the velocity-averaged 
$\langle \sigma _{T}\rangle$ and non-perturbative effects when $\alpha_\phi M_X>M_{H_2}$.

As an illustration, we show the scatter plots for $M_{H_2}$-$M_X$. 
Since we focus on the light $H_2$ here, we can fix $M_{Z'}=200$ GeV and impose 
the constrain from electroweak precison observable, $\epsilon \ll 0.03$. 
Other parameters are scanned as indicated from the legend bar of individual plot. 
\[
g_X\lesssim 1.2,\; \lambda_{\phi X}\lesssim 1,\; \lambda_{H X}\lesssim 0.1,\; \lambda_{3}\lesssim 0.1,\;\textrm{and } \lambda_{\phi H}\simeq 0.
\]
Because of the velocity-dependent behavior of Eq.~\ref{eq:SIcs} and \ref{eq:transfercs}, 
the transfer cross section over mass, $\sigma_T/M_X$, can be around $\left[0.1,10\right]\textrm{ cm}^{2}/\textrm{g}$ at Dwarf scale with $v_\textrm{rel}\simeq 10$ km/s while still 
satisfy the requirement $\sigma_T/M_X\lesssim 0.5\textrm{ cm}^{2}/\textrm{g}$ 
to be consistent with ellipticity constraints on Milky Way and cluster scales. 

Before closing this section, we briefly discuss the CMB constraints which are quite strong. 
When $\alpha_{\phi}M_X>M_\phi$, there would exist large non-perturbative effect in the 
low-velocity limit ($v\rightarrow 0$) of DM particle, known as Sommerfeld enhancement,  
and there could be relevant astrophysical constraint from cosmic microwave 
background(CMB) for some parameter space we discussed above. 
Then $XX^*$ annihilation is enhanced at CMB time and significant energy would be 
injected to photon-baryon bath, broadening the last scattering surface and leaving an imprint 
in CMB spectra~\cite{Galli:2009zc, Slatyer:2009yq,Cirelli:2009bb, Hutsi:2011vx, Natarajan:2012ry, 
Cline:fm, Diamanti:2013bia, Madhavacheril:2013cna}. 
Current data constrains the enhancement factor $S\lesssim\mathcal{O}(1000)$ for 
$\mathcal{O}(\TeV)$ DM with the 
exact value depending on the specific annihilation channel. As an illustration, taking 
parameters for large self-interactions for the DM's such as 
\[
M_X\simeq 1\TeV,\; M_\phi \simeq 1\MeV,\; \lambda_{\phi X}\simeq 0.1,
\]
we find that the enhancement factor saturates at $S\sim \mathcal{O}(50)$, 
which is well below the current limit, $S\lesssim\mathcal{O}(1000)$.  
Therefore the discussions on self-interacting DM presented in this section are safe from  
the CMB constraints. 

Finally let us add that this mechanism for enhancing the DM self-interactions could be
realized not only by light scalar mediator $\phi$ but also by a light vector mediator $Z^{'}$ 
between $X$ and $X^*$ (namely, between opposite dark charges). 
Thus this feature is not unique to local $Z_3$ models, and could be easily realized in 
other models too, such as local $Z_2$ models~\cite{preparation}.

\section{Summary}

In this paper, we have proposed a self-interacting scalar DM model with a local dark $Z_3$ 
symmetry. Unlike global dark symmetries, local ones can guarantee that DM is absolutely 
stable even in the presence of higher dimensional nonrenormalizable operators due to 
the underlying local gauge symmetry.   Then we discussed perturbativity constraints 
on the scalar potential and the experimental limit on the kinetic mixing. 
Compared with a global $Z_3$ model, our scenario has two new particles, $Z^{'}$ and 
$H_2$,  and there are new channels in the DM pair annihilations for thermalizing DMs. 
Therefore much ampler parameter space is allowed including a light DM with 
$M_X<125$ GeV,  most region of which can be probed with future DM direct searches. 
Also, motivated by the small scale astrophysical anomalies, we investigated 
the phenomenology of a MeV scalar $H_2$ in our model which has no counterpart in 
the minimal global $Z_3$ model. Thanks to the velocity dependence of DM self-interaction 
cross section,  such a light $H_2$ can mediate strong interaction for DM scattering at 
Dwarf galaxy scale while satisfying Milky Way and cluster scale constraints. Similar arguments go for the light $Z^{'}$ as well.  For such a light $H_2$ or $Z^{'}$, there could be exotic decays of the 126 GeV Higgs boson, which could be studied in the upcoming LHC running and at future lepton colliders.

\begin{acknowledgments}
We are grateful to Seungwon Baek and Wan-Il Park for useful comments and discussions. 
This work is supported in part by National Research Foundation of Korea (NRF) Research 
Grant 2012R1A2A1A01006053 (PK,YT), and by the NRF grant funded by the Korea 
government (MSIP) (No. 2009-0083526) through  Korea Neutrino Research Center 
at Seoul National University (PK).
\end{acknowledgments}

\section{Appendix}

\subsection{RGEs}

For future reference, here we present the RGEs in the case of no kinetic mixing, 
\begin{align*}
\frac{d\lambda_{H}}{d\ln\mu} & =\frac{1}{16\pi^{2}}\left[24\lambda_{H}^{2}+\lambda_{\phi H}^{2}+\lambda_{HX}^{2}-6y_{t}^{4}+\frac{3}{8}\left(2g_{2}^{4}+\left(g_{1}^{2}+g_{2}^{2}\right)^{2}\right)-\lambda_{H}\left(9g_{2}^{2}+3g_{1}^{2}-12y_{t}^{2}\right)\right],\\
\frac{d\lambda_{\phi}}{d\ln\mu} & =\frac{1}{16\pi^{2}}\left[20\lambda_{\phi}^{2}+2\lambda_{\phi H}^{2}+\lambda_{\phi X}^{2}+6g_{X}^{4}-12\lambda_{\phi}g_{X}^{2}\right],\\
\frac{d\lambda_{X}}{d\ln\mu} & =\frac{1}{16\pi^{2}}\left[20\lambda_{X}^{2}+2\lambda_{HX}^{2}+\lambda_{\phi X}^{2}+9\lambda_{3}^{2}+\frac{2}{27}g_{X}^{4}-\frac{4}{3}\lambda_{\phi}g_{X}^{2}\right],\\
\frac{d\lambda_{\phi H}}{d\ln\mu} & =\frac{1}{16\pi^{2}}\left[4\lambda_{\phi H}\left(3\lambda_{H}+2\lambda_{\phi}+\lambda_{\phi H}\right)+\lambda_{\phi X}\lambda_{HX}-\lambda_{\phi H}\left(\frac{9}{2}g_{2}^{2}+\frac{3}{2}g_{1}^{2}-6y_{t}^{2}+6g_{X}^{2}\right)\right],\\
\frac{d\lambda_{HX}}{d\ln\mu} & =\frac{1}{16\pi^{2}}\left[4\lambda_{HX}\left(3\lambda_{H}+2\lambda_{X}+\lambda_{HX}\right)+\lambda_{\phi H}\lambda_{\phi X}-\lambda_{HX}\left(\frac{9}{2}g_{2}^{2}+\frac{3}{2}g_{1}^{2}-6y_{t}^{2}+\frac{2}{3}g_{X}^{2}\right)\right],\\
\frac{d\lambda_{\phi X}}{d\ln\mu} & =\frac{1}{16\pi^{2}}\left[2\lambda_{\phi X}\left(2\lambda_{\phi}+2\lambda_{X}+\lambda_{\phi X}\right)+2\lambda_{\phi H}\lambda_{HX}+18\lambda_{3}^{2}-\lambda_{HX}\left(6g_{X}^{2}+\frac{2}{3}g_{X}^{2}\right)\right],\\
\frac{dg_{X}}{d\ln\mu} & =\frac{1}{16\pi^{2}}\left(\frac{1}{3}+\frac{1}{27}\right)g_{X}^{3},\\
\frac{d\lambda_{3}}{d\ln\mu} & =\frac{1}{16\pi^{2}}\left[\lambda_{3}\left(2\lambda_{X}+\lambda_{\phi X}\right)\right].
\end{align*}

\subsection{Positive Conditions for Quartic Polynomial}

This section summarizes the positivity conditions for quartic polynomials,
see Ref.\cite{polynomial} for mathematical details. For a general
quartic polynomial 
\begin{equation}
f(z)=az^{4}+bz^{3}+cz^{2}+dz+e,
\end{equation}
with real coefficients, positive $a$ and $e$, $f\left(z\right)\geq0$
for $z>0$ shall constrain the regions of coefficients. Positivity
on any fixed interval $\left(u,v\right)$ can be translated directly
to positivity on the positive reals through the transformation 
\[
t=\frac{u+zv}{1+z}.
\]
With the replacement $x^{4}=\frac{a}{e}z^{4},$ the polynomial $f\left(z\right)/e$
then becomes $p\left(x\right)=x^{4}+\alpha x^{3}+\beta x^{2}+\gamma x+1,$
where we have defined
\[
\alpha=ba^{-\frac{3}{4}}e^{-\frac{1}{4}},\beta=ca^{-\frac{1}{2}}e^{-\frac{1}{2}},\gamma=da^{-\frac{1}{4}}e^{-\frac{3}{4}}.
\]
Now the question is shifted to the positivity of $p\left(x\right)$
for $x\geq0$. Define
\begin{eqnarray}
 &  & \Delta=4\left[\beta^{2}-3\alpha\gamma+12\right]^{3}-\left[72\beta+9\alpha\beta\gamma-2\beta^{3}-27\alpha^{2}-27\gamma^{2}\right]^{2},\\
 &  & \Lambda_{1}\equiv(\alpha-\gamma)^{2}-16(\alpha+\beta+\gamma+2),\;\\
 &  & \Lambda_{2}\equiv(\alpha-\gamma)^{2}-\frac{4(\beta+2)}{\sqrt{\beta-2}}\left(\alpha+\gamma+4\sqrt{\beta-2}\right).
\end{eqnarray}
Then $p\left(x\right)\geq0$ for all $x\geq0$ or $f(z)\geq0$ for
all $z>0$ if and only if
\begin{align}
(1)\; & \beta<-2\;\mathrm{and}\;\Delta\leq0\;\mathrm{and}\;\alpha+\gamma>0;\\
(2)\; & -2\leq\beta\leq6\textrm{ and }\begin{cases}
\Delta\leq0 & \textrm{and }\alpha+\gamma>0\\
\Delta\geq0 & \textrm{and }\Lambda_{1}\leq0;
\end{cases}\\
(3)\; & 6<\beta\textrm{ and }\begin{cases}
\Delta\leq0 & \textrm{and }\alpha+\gamma>0\\
\alpha>0 & \textrm{and }\gamma>0\\
\Delta\geq0 & \textrm{and }\Lambda_{2}\leq0.
\end{cases}
\end{align}
It is also useful to give the following sufficient conditions for
positivity,
\begin{align}
(1)\; & \alpha>-\frac{\beta+2}{2}\;\mathrm{and}\;\gamma>-\frac{\beta+2}{2}\;\textrm{ for }\beta\leq6,\\
(2)\; & \alpha>-2\sqrt{\beta-2}\;\mathrm{and}\;\gamma>-2\sqrt{\beta-2}\;\textrm{ for }\beta>6.
\end{align}

\end{document}